\documentclass[10pt,aps,prd,reprint,superscriptaddress,nofootinbib,preprintnumbers]{revtex4-1}

\usepackage[utf8]{inputenc}
\usepackage{amsmath}
\usepackage{amssymb}
\usepackage{fontawesome}
\usepackage{color}
\usepackage{listings}
\usepackage{graphicx}
\usepackage{multirow}
\usepackage{caption}
\usepackage[normalem]{ulem}
\usepackage[colorlinks,pdfusetitle]{hyperref}
\hypersetup{colorlinks=true,allcolors=[rgb]{1,0.56,0}}

\definecolor{dkgreen}{rgb}{0,0.6,0}
\definecolor{gray}{rgb}{0.5,0.5,0.5}
\definecolor{mauve}{rgb}{0.58,0,0.82}
\newcommand{\orcid}[1]{\href{https://orcid.org/#1}{#1}}
\newcommand{\e}[1]{\times10^{#1}}

\newcommand{\lam}[1]{\lambda_{#1}}
\newcommand{\V}[1]{V_{#1}}

\begin{document}

\title{Fast and Accurate Algorithm for Calculating Long-Baseline Neutrino Oscillation Probabilities with Matter Effects: NuFast}

\author{Peter B.~Denton}
\email{peterbd1@gmail.com}
\thanks{\orcid{0000-0002-5209-872X}}
\affiliation{High Energy Theory Group, Physics Department, Brookhaven National Laboratory, Upton, NY 11973, USA}

\author{Stephen J.~Parke}
\email{parke@fnal.gov}
\thanks{\orcid{0000-0003-2028-6782}}
\affiliation{Theoretical Physics Department, Fermi National Accelerator Laboratory, Batavia, IL 60510, USA}

\preprint{Fermilab-Pub-24-0207-T}

\date{May 3, 2024}

\begin{abstract}
Neutrino oscillation experiments will be entering the precision era in the next decade with the advent of high statistics experiments like DUNE, HK, and JUNO.
Correctly estimating the confidence intervals from data for the oscillation parameters requires very large Monte Carlo data sets involving calculating the oscillation probabilities in matter many, many times.
In this paper, we leverage past work to present a new, fast, precise technique for calculating neutrino oscillation probabilities in matter optimized for long-baseline neutrino oscillations in the Earth's crust including both accelerator and reactor experiments.
For ease of use by theorists and experimentalists, we provide fast c++ and fortran codes.
{\href{https://github.com/PeterDenton/NuFast}{\large\faGithub}}
\end{abstract}

\maketitle

\section{Introduction}
Neutrino oscillation is the curious process by which sources of one flavor of neutrino are sometimes detected later as a different flavor.
This phenomenon provides evidence for physics beyond our Standard Model of particle physics.
Understanding the nature of neutrinos and the fundamental parameters that govern neutrino oscillations is the central focus of a massive international experimental effort comprised of an alphabet soup of experiments.

Thus far, a rough picture of many of the oscillation parameters has become clear \cite{deSalas:2020pgw,Capozzi:2021fjo,Esteban:2020cvm,Denton:2022een}.
The magnitudes of both mass squared differences are known and the sign of one of them is known.
The general size of three mixing angles are known.
What remains to be determined about neutrinos via oscillations is the sign of the other mass squared difference, the octant of one mixing angle, and the value of the complex phase.
In addition, measurements of the so-called ``solar parameters'' in coming years will enter the precision era.

The current generation long-baseline (LBL) accelerator neutrino oscillation experiments, NOvA \cite{Ayres:2004js} and T2K \cite{Itow:2001ee} are making some progress on these parameters and will continue to collect data for several years.
As we progress to the second generation LBL accelerator experiments, DUNE \cite{Abi:2020evt} and HK \cite{Abe:2018uyc} as well as the LBL reactor experiment JUNO \cite{JUNO:2021vlw},
we will collect enough information to answer all three questions.
As the field moves to the precision era with the determinations of parameters at the sub-percent level and possible determinations of parameters at the $>5\sigma$ level, several issues that could previously be ignored must now be handled more carefully.
For example, estimating the preferred value of a given parameter from a data set requires careful statistical analysis, often via Monte-Carlo estimation, see e.g.~\cite{Feldman:1997qc} (Feldman and Cousins).
This requires performing many throws of an experiment at each point in parameter space.
A careful examination of computational efforts has found that in some such analyses, the majority of the computer time spent is on many calculations of the neutrino oscillation probabilities \cite{novaprivate}.
For example, for a single oscillation analysis, NOvA used some 35 million core hours \cite{nova_computing:2018}.
This issue will only be exacerbated with next generation experiments DUNE, HK, and JUNO which aim for significant increases in precision requiring additional Monte Carlo throws.
Thus any improvement in the neutrino oscillation probability calculation will reduce computational costs.

The role of the matter effect on the appearance of a new neutrino flavor via oscillations is a fundamentally three-flavor problem which is non-trivial to approximate well in matter \cite{Barenboim:2019pfp}.
In addition, while additional insights can be gained about disappearance in matter in a two-flavor picture for $\nu_e$ disappearance \cite{Denton:2018cpu} and $\nu_\mu$ disappearance \cite{Denton:2024thm},
more accuracy is likely required by next generation experiments.
Here we present an algorithm that handles both appearance and disappearance channels in matter in a unified framework that is efficient and easily significantly more precise than needed by next generation experiments.

In this work, we will review the problem in section \ref{sec:problem}.
Building on work understanding neutrino oscillations in matter, along with some new techniques presented here for the first time in this context,
we will present our optimal algorithm for computing neutrino oscillations in matter for LBL accelerator or reactor experiments in section \ref{sec:algorithm}.
In section \ref{sec:code} we detail how the publicly available code, \verb1NuFast1, works in both fortran and c++.
We will quantify the performance of this approach in section \ref{sec:performance} by comparing it to the exact analytic expression for numerical precision across a range of kinematic and fundamental parameters relevant for these experiments.
We will also quantify the speed of it relative to existing expressions on the market.
Finally, we conclude in section \ref{sec:conclusions}.

\section{The Problem}
\label{sec:problem}
Neutrino oscillation experiments aim to measure six fundamental parameters: $\Delta m^2_{31}$, $\Delta m^2_{21}$, $\theta_{12}$, $\theta_{13}$, $\theta_{23}$, and $\delta$.
For experiments in vacuum (or in the Earth's atmosphere) it is straightforward to relate what is measured, the neutrino oscillation probabilities, to these six underlying parameters at a given distance traveled $L$ and neutrino energy $E$.
The probabilities describe the ratio of actually detected neutrinos to the amount that would be detected if neutrinos did not oscillate.
If the neutrinos propagate in dense environments, such as the Earth or the Sun, the presence of matter must be accounted for \cite{Wolfenstein:1977ue}.
While there are numerous techniques to write the oscillation probabilities in matter as functions of the underlying parameters, in this article we aim to do so leveraging recent theoretical work and targeting approximations relevant given the oscillation parameters that have been well measured and are small such as $\theta_{13}$ and $\Delta m^2_{21}/|\Delta m^2_{31}|$.

In vacuum, the neutrino oscillations are governed by a simple Hamiltonian, given here in the flavor basis as
\begin{equation}
H_{\rm vac}=\frac1{2E}U
\begin{pmatrix}
0\\&\Delta m^2_{21}\\&&\Delta m^2_{31}
\end{pmatrix}U^\dagger\,,
\end{equation}
where $U$ is the PMNS matrix \cite{Pontecorvo:1957cp,Maki:1962mu}.
It can be parameterized in many different ways, see e.g.~\cite{Denton:2020igp}, including the popular approach described in the PDG \cite{ParticleDataGroup:2022pth} as
\begin{equation}
U\hspace{-1mm}=\hspace{-1mm}
\begin{pmatrix}
1\\
&c_{23}&s_{23}\\
&-s_{23}&c_{23}
\end{pmatrix}
\hspace{-2mm}
\begin{pmatrix}
c_{13}&&s_{13}e^{-i\delta}\\
&1\\
-s_{13}e^{i\delta}&&c_{13}
\end{pmatrix}
\hspace{-2mm}
\begin{pmatrix}
c_{12}&s_{12}\\
-s_{12}&c_{12}\\
&&1
\end{pmatrix}
\hspace{-1mm},
\end{equation}
where $s_{ij}=\sin\theta_{ij}$ and $c_{ij}=\cos\theta_{ij}$.

Given the Hamiltonian, there are numerous ways to calculate the probability.
One conceptually simple way is to recall that solutions to the Schr\"odinger equation are exponentials of the Hamiltonian times the time.
Equating the time with the distance gives
\begin{equation}
\mathcal A=\exp(-iHL)\,,
\label{eq:amplitude}
\end{equation}
where $\mathcal A$ is the matrix of amplitudes of each of the nine oscillation probability channels.
The probabilities are then given by
\begin{equation}
P(\nu_\alpha\to\nu_\beta)=|\mathcal A_{\alpha\beta}|^2\,.
\end{equation}
Evaluating eq.~\ref{eq:amplitude} requires diagonalizing the Hamiltonian.
That is, one must determine both the eigenvalues\footnote{Note that while $\lambda_i/2E$ are actually the eigenvalues, for convenience,
we sometimes call the $\lambda_i$, without the $1/2E$ factor, as the eigenvalues.} ($\lambda_1$, $\lambda_2$, and $\lambda_3$)/2E and the eigenvectors which form into the matrix $V$.
By the properties of exponentials, eq.~\ref{eq:amplitude} can then be written as
\begin{equation}
\mathcal A=V
\begin{pmatrix}
e^{-i\lambda_1L/2E}\\&e^{-i\lambda_2L/2E}\\&&e^{-i\lambda_3L/2E}
\end{pmatrix}V^\dagger\,,
\end{equation}
which is trivial to calculate, once the eigenvalues and eigenvectors have been determined.\\

It is easy to see that, in vacuum, the eigenvalues and eigenvectors follow directly from the vacuum expressions.
In matter, however, the situation is more complicated.
This is because when neutrinos traverse a sufficiently dense electron field, such as in the Earth or the Sun, the Hamiltonian is significantly modified by the addition of a term,
\begin{equation}
H=H_{\rm vac}+\frac1{2E}
\begin{pmatrix}
a\\&0\\&&0
\end{pmatrix}\,,
\end{equation}
where $a$ quantifies the size of the Wolfenstein matter effect \cite{Wolfenstein:1977ue} and is
\begin{align}
a&=2\sqrt2G_FN_eE\,,\\
&=1.52\e{-4}\left(\frac{Y_e\rho}{\rm g/\rm{cm}^3}\right)\left(\frac E{\rm{GeV}}\right)\rm{eV}^2\,,\nonumber
\end{align}
and $G_F$ is Fermi's constant, $N_e$ is the electron density, $Y_e$ is the electron fraction and is $\sim0.5$ in the Earth, and $\rho$ is the density of the matter.

In the presence of matter, it is now nontrivial to diagonalize the Hamiltonian when matter effects matter, which they do for long-baseline accelerator and reactor experiments (see \cite{Li:2016txk,Khan:2019doq}), as well as for atmospheric and solar experiments.

The next section describes our means of diagonalizing the Hamiltonian, focused on realistic oscillation parameters and computational speed of the method
given that the experiments will make an enormous number of oscillation probability evaluations when using the Feldman-Cousins technique.

\section{The Algorithm}
\label{sec:algorithm}
The algorithm described below, in essence, consists of two layers, each with some innovation.
The first is the application of an approximation for $\lambda_3$ using properties of the characteristic equation to get the eigenvalues, combined with the adjugate method to get the eigenvectors.
These are then combined to get all nine oscillation probabilities.
The second layer is that, simultaneously with the steps just described, we make choices to minimize the number of expensive function calls.
After performing numerous computational tests on different computers and in different programming languages, we consistently find that trig functions (sine and cosine) are very slow; square roots are also quite slow, although not as bad as trig functions. Next slowest is the division operation and therefore their number should be minimize.
Other operations are significantly faster and are essentially irrelevant in comparison.
We also reduce multiplications in favor of additions when possible.
The exact computational speed of each basic operation will depend on the architecture, language, and what optimizations the compiler chooses to take.
Nonetheless, our optimization conditions seem to be fastest for a wide variety of scenarios.
Most of the details are described here and the exact implementations are in the code.

\subsection{Vacuum}
The algorithm we give here starts with the oscillation probabilities in vacuum and then we modify the relevant variables, from their vacuum values, to the equivalent variables in matter.
We assume that the input neutrino parameters are given by: $\Delta m^2_{21}, ~\Delta m^2_{31}, ~s^2_{23}, ~s^2_{13}, ~s^2_{12}, ~ \delta$.
A different choice of inputs may slow things down a bit, such as using the mixing angles, $\theta_{ij}$, due to the necessity of evaluating three additional trig functions.
The experimental input parameters are the neutrino energy, baseline, and for propagation in matter the average matter density between source and detector, as well as the electron fraction (typically $\sim0.5$ in the Earth): $E$, $L$, $\rho$, and $Y_e$, respectively.

First we calculate the squares of elements of the PMNS matrix, $U$, and the Jarlskog invariant \cite{Jarlskog:1985ht}, $J$, as 
\begin{align}
& |U_{e2}|^2 =c^2_{13}s^2_{12}\,, \quad |U_{e3}|^2=s^2_{13}\,, \quad
 |U_{\mu 3}|^2 =c^2_{13}s^2_{23}\,, \notag \\
& |U_{\mu2}|^2 =c^2_{12}c^2_{23}+ s^2_{13}s^2_{12}s^2_{23}-2 J_{rr} \cos \delta\,,\\
&J = J_{rr} c^2_{13}\sin \delta\,, ~~ \text{with} ~~J_{rr} =\sqrt{c^2_{12}c^2_{23} s^2_{13}s^2_{12}s^2_{23}} \,,
\notag 
\end{align}
where $c^2_{ij} \equiv 1-s^2_{ij}$.
All other $ |U_{\alpha i}|^2$'s can by calculated by unitarity, using
$|U_{\alpha 1}|^2=1-|U_{\alpha 3}|^2-|U_{\alpha 2}|^2$ or $|U_{\tau i}|^2=1-|U_{\mu i}|^2-|U_{e i}|^2$, 
as needed.

Then the disappearance and appearance vacuum oscillation probabilities, using the kinematic phase $\Delta_{ij}=\Delta m^2_{ij}L/4E$, are given by 
\begin{align}
P_{\alpha \alpha} ={}&1 -4\sum_{i>j} |U_{\alpha i}|^2 |U_{\alpha j}|^2 \sin^2 (\Delta_{ij})\,, \\
P_{\alpha \beta} ={}& -4 \sum_{i>j} R^{ij}_{\alpha \beta} \sin^2 ( \Delta_{ij} ) \notag \\
&-8 J_{\alpha \beta} \sin ( \Delta_{21} ) \sin ( \Delta_{31} ) \sin ( \Delta_{32} )\,,
\label{eq:oscprob}
\end{align}
where $\alpha \neq \beta$, $J_{\mu e}=J_{e \tau}=J_{\tau \mu}= J$ and $J_{\beta \alpha}=-J_{\alpha \beta}$. Also
\begin{align}
R^{ij}_{\alpha \beta} &\equiv\Re(U_{\alpha i} U^*_{\beta j} U^*_{\alpha j}U_{\beta i}) \\
& =\frac1{2}( |U_{\alpha k}|^2 |U_{\beta k}|^2- |U_{\alpha i}|^2 |U_{\beta i}|^2- |U_{\alpha j}|^2 |U_{\beta j}|^2 )\,, \notag 
\end{align}
where $k$ $\neq$ $i$, $j$.
This identity\footnote{\label{ftnote:RJid}A useful consistency condition here is that, for all $\alpha \neq \beta$ and $i \neq j $,
$(R^{ij}_{\alpha \beta})^2=|U_{\alpha i}|^2 |U_{\beta j}|^2 |U_{\alpha j}|^2 |U_{\beta i}|^2 - J^2$.}
for $R^{ij}_{\alpha \beta}$ is not commonly used and is only valid for 3 flavors. It comes from squaring the unitarity triangle closure: $U_{\alpha i} U^*_{\beta i}+U_{\alpha j} U^*_{\beta j}=-U_{\alpha k} U^*_{\beta k} $.
 
One only needs to calculate two disappearance probabilities and one appearance probability, provided that the appearance probability is split into CP-conserving and CP-violating parts.
That is, the following are sufficient to compute all nine channels: $P_{ee}$, $P_{\mu \mu}$, $P^{CPC}_{\mu e}$, and $P^{CPV}_{\mu e}$.

For the calculation of the vacuum oscillation probabilities, we have only used 
\begin{equation*}
(|U_{\alpha i}|^2, ~ J, ~\Delta m^2_{ij})\,.
\end{equation*}
The calculation of the three sines of the kinematic phases, $\sin ( \Delta_{ij} )$, together with $\sin \delta$ and $\cos \delta$, uses on the order of 50\% of the computational
effort for the calculation of the vacuum oscillation probabilities with this algorithm.
Note that if one computes the amplitude first and then takes the norm-squared, this requires computing four trig functions of kinematic phases (the real and complex part for each of the two frequencies) instead of just three.
This gives us confidence that the approach described here for the general form of the neutrino oscillation probabilities is the most efficient.
If the form of the inputs parameters are changed (e.g.~to the mixing angles instead of $s_{ij}^2$), this algorithm will have to be slightly modified to accommodate this change.

\subsection{Matter}
To calculate the oscillation probabilities in constant matter, we need the equivalent of $(|U_{\alpha i}|^2, ~ J, ~\Delta m^2_{ij})$ in matter.
To do this we have to find the solutions to the characteristic equation and then calculate particular combinations of the elements of the eigenvectors.
Here, we have developed a method to do this which is computationally efficient and accurate.
What is required is combinations of the trace, the determinant, and the diagonal elements of the adjugate of the Hamiltonian in the flavor basis
to determine the coefficients in both the characteristic equation and the Eigenvector-Eigenvalue identity \cite{Denton:2019ovn,Denton:2019pka} as given below.

First, for the characteristic equation, which only depends on $\Delta m^2_{31},~\Delta m^2_{21}$, $|U_{e3}|^2$, $|U_{e2}|^2$ and the Wolfenstein matter potential times $2E$, given by $a \equiv 2\sqrt{2} G_F N_e E$, we have
\begin{align}
A={}&(2E) \,\text{Tr}[H]=\Delta m^2_{21}+\Delta m^2_{31}+a\,,\\[2mm]
B ={}&(2E)^2 \,\frac{1}{2}(\text{Tr}^2[H]-\text{Tr}[H^2]) =\Delta m^2_{21}\Delta m^2_{31} \notag \\
& +a[\Delta m^2_{21} (1-|U_{e2}|^2)+\Delta m^2_{31} (1-|U_{e3}|^2)] \,, \\[2mm]
C={}&(2E)^3 \,\text{Det}[H]=a \Delta m^2_{21}\Delta m^2_{31}|U_{e1}|^2 \,.
\end{align}
 Then to find the eigenvalues of the Hamiltonian for the neutrinos propagating in matter we have to solve the characteristic equation:
 \begin{align}
 X(\lam{ })&=\lam{ }^3-A \lam{ }^2+B\lam{ }-C=0 \,.
 \label{eq:charX}
 \end{align}
The exact solution\footnote{The exact $\lam{3}$ is given in \cite{cardano,Barger:1980tf, Zaglauer:1988gz}, as
\begin{align}
\lam{3}={}&\frac1{3}A+\frac2{3} \sqrt{A^2-3B} \notag \\
& \times \cos\left[\frac1{3}\left(\arccos\left[\frac{2A^3-9AB+27C}{2 \left( \sqrt{A^2-3B}\,\right)^3}\right]+2\pi n \right) \right] 
\end{align}
with $n=0$ for NO and $n=1$ for IO.} for this cubic equation is computationally very expensive as it involves the calculation
of $\cos[(\arccos[\cdots]+2\pi n)/3]$, so we use a simple but accurate approximate solution, given by
Denton, Minakata and Parke (DMP) \cite{Minakata:2015gra,Denton:2016wmg}:
\begin{align}
\lam{3}={} &\Delta m^2_{31} +\frac1{2}\Delta m^2_{ee}\left(x-1+\sqrt{(1-x)^2+4 x s^2_{13} } ~\right)\,,
\notag \\
\text{with} ~x&\equiv \frac{a}{\Delta m^2_{ee}} ~ \text{and} ~ \Delta m^2_{ee} \equiv \Delta m^2_{31}-s^2_{12}\,\Delta m^2_{21} \,.
\label{eq:DMP+}
\end{align}
$\Delta m^2_{ee}$ is the $\Delta m^2$ measured in $\nu_e$ disappearance experiments as defined in \cite{Nunokawa:2005nx}, see also \cite{Parke:2016joa}.
This approximation for $\lam{3}$ is an excellent one for both mass orderings 
with a fractional difference to the exact eigenvalue of better than $10^{-4}$ for all neutrino energies, and is exact in vacuum\footnote{The
 argument of the square root, in eq.~\ref{eq:DMP+}, is non-negative for all $x$, thus $\lam{3}$ is always real.
 Also, since $|\lam{3}| > c^2_{13} |\Delta m^2_{31}|$, it is never zero and therefore can be used in the denominator of eq.~\ref{eq:Dlambda21}, when calculating the other eigenvalues.}.
 
However, if needed, it can be further improved by using one or more Newton-Raphson (NR) iterations\footnote{Other
improvement methods could be used here, such as 2nd order NR, however we use 1st order NR because of its simplicity and rapid rate of convergence.}:
\begin{equation}
\lam{3} \rightarrow \lam{3} - \frac{X(\lam{3}) }{ X^\prime(\lam{3}) }\,.
\label{eq:NR}
\end{equation}
Here, $X^\prime$ is the derivative of the characteristic equation $X(\lambda)$ (eq.~\ref{eq:charX}) with respect to $\lambda$.

The improvement in the accuracy for $\lam{3}$ with these iterations is quadratic.
The convergence rate of the NR iterations is quite sensitive to how close the initial approximation is to the exact solution.
Given that eq.~\ref{eq:DMP+} is an excellent approximation for the exact eigenvalue as discussed in quantitative detail in the appendix,
it is not surprising that the NR iterations lead to a very rapid convergence to the exact eigenvalue, and thus the exact probability,
at the level of about 5 orders of magnitude improvement on the probability for the first iteration, and even faster thereafter.

The remaining two eigenvalues can then be easily obtained by solving two equations: we have chosen to use A, C for simplicity, but using A, B or B, C give similar results:
\begin{align}
\lam{1}+\lam{2} &= A-\lam{3}\,, \quad \lam{1}\lam{2} = C/\lam{3}\,, \notag\\
\text{as} ~ \Delta \lam{21}&=\sqrt{( A-\lam{3} )^2-4C/\lam{3}}\,.
\label{eq:Dlambda21}
\end{align}
The sign of the square root is the same as the sign of $\Delta m^2_{21}$ which is known to be positive from solar neutrino data.
This then gives you $\lam{2,1}=\frac1{2}(A-\lam{3} \pm \Delta \lam{21})$ with $\lam{2}>\lam{1}$.

Second, for the coefficients of the Eigenvector-Eigenvalue identity we also need $|U_{\mu 2}|^2$ and $|U_{\mu 3}|^2$, see \cite{Abdullahi:2022fkh}.
The coefficients for the identity are given by\footnote{The diagonal elements of the Hamiltonian in the flavor basis, $H_{\alpha \alpha}= \Delta m^2_{21} |U_{\alpha 2}|^2+\Delta m^2_{31} |U_{\alpha 3}|^2+a\delta_{e \alpha}$, appears frequently in B, $ S_{\alpha \alpha}$ and $T_{\alpha \alpha}$ and could be used to make these expressions appear simpler. The off-diagonal elements of H do not appear here. }
\begin{align}
S_{\alpha \alpha} &= (2E)\,(\text{Tr}[H] I -H)_{\alpha \alpha}\\
&= \Delta m^2_{21} (1-|U_{\alpha 2}|^2)+\Delta m^2_{31}(1-|U_{\alpha 3}|^2)+a(1-\delta_{\alpha e}) \notag
\end{align}
\begin{align}
T_{\alpha \alpha} ={}& (2E)^2 \,  \text{Adjg}[H]_{\alpha \alpha} 
=	\Delta m^2_{21}\Delta m^2_{31} |U_{\alpha 1}|^2\notag\\
&+ a(1-\delta_{\alpha e}) (\Delta m^2_{21} |U_{\beta 2}|^2+\Delta m^2_{31} |U_{\beta 3}|^2 ) \, ,
\end{align}
where $\beta \neq e$ or $\alpha$.
Note these are all simple functions of the $\Delta m^2_{ij}$'s, $|U_{\alpha i}|^2$'s and ``a''. The effects of 
$\sin ^2\theta_{23}$ and $\cos \delta$ enter here, through $|U_{\mu 3}|^2$ and $|U_{\mu 2}|^2$.

Calculation of all of these coefficients, A, B, C, $S_{\alpha \alpha}$ and $T_{\alpha \alpha}$, can be fully automated using the Le\,Verrier-Faddeev algorithm, and are all related, see section III-2 of \cite{Abdullahi:2022fkh}. In particularly, they must satisfy the following relationships:
$S_{ee}+S_{\mu \mu} + S_{\tau \tau} = 2A$ and $T_{ee}+T_{\mu \mu} + T_{\tau \tau} = B$.

With the these coefficients and the eigenvalues in place, we can use the Eigenvector-Eigenvalue identity to obtain the matter equivalents of $|U_{\alpha i}|^2$:
\begin{align}
| \V{\alpha i} |^2 &= \frac{\lam{i}^2-S_{\alpha \alpha} \lam{i}+ T_{\alpha \alpha}}
{\Delta \lam{ij} \Delta \lam{ik} }\,,
\label{eq:EE}
\end{align}
where $j$ and $k$ are not equal to $i$ or each other.
As these squared elements of the eigenvectors are properly normalized
\begin{equation}
\sum_i | \V{\alpha i} |^2=\sum_\alpha | \V{\alpha i} |^2=1\,,
\end{equation}
only $| \V{e 3} |^2,| \V{e 2} |^2$ and $| \V{\mu 3} |^2,| \V{\mu 2} |^2$ need to be calculated using eq.~\ref{eq:EE}, as the others can be obtained via unitarity.

At this stage the only thing left to calculate before we can use eq.~\ref{eq:oscprob} for the matter oscillation probabilities is the Jarlskog invariant in matter. For this we use the Naumov-Harrison-Scott (NHS) identity \cite{Naumov:1991ju, Harrison:2002ee}:
\begin{align}
J_{mat} = J ~ \Pi_{i>j} \left( \frac{\Delta m^2_{ij}}{\Delta \lam{ij}} \right) \,.
\end{align}
The effects of $\sin \delta$ enter here, as this is the CP violating term.
Again one can use consistency check given in the footnote \ref{ftnote:RJid} see section \ref{sec:algorithm} for $|\V{\alpha i}|^2$ and $J_{mat}$.

So with the replacement of $$(|U_{\alpha i}|^2, ~ J, ~ \Delta m^2_{ij}) ~~ \Rightarrow ~~ (|\V{\alpha i}|^2, ~J_{mat}, ~\Delta \lam{ij})$$ in eq.~\ref{eq:oscprob} we can now calculate all the oscillation probabilities in matter.
This algorithm is set up such that the neutrino (anti-neutrino) oscillation probabilities are given with $E > 0$ ($ < 0$), the mass ordering is determined by the sign of $\Delta m^2_{31}$: NO (IO) is $\Delta m^2_{31}>0$ ($<$ 0),
and propagation in matter (anti-matter) is given by positive (negative) medium density, $\rho$.
Switching the sign of the baseline, L, changes $\nu_\alpha \rightarrow \nu_\beta$ to $\nu_\beta \rightarrow \nu_\alpha$, but as the algorithm returns all nine oscillation probabilities this is not needed.

The designed usage of this algorithm is for the long-baseline experiments T2K/HK, NOvA, and DUNE and is also valid for JUNO and has been carefully tested within the scope of these experiments.
If the parameters are significantly varied beyond the standard three-flavor oscillation picture by many $\sigma$ away from the current best fit values, the NR iterations may not converge as rapidly as expected.
Our aim here is speed rather than broad applicability to any conceivable neutrino oscillation problem and we focused on the above experiments.

The accuracy and efficiency of this algorithm will be discussed in the next section.

\section{Code}
\label{sec:code}
There are numerous approaches through which one could organize the code, each with its own advantages and disadvantages.
We make two key assumptions at the outset.
First, we assume that all nine oscillation channels are of interest for a given problem.
For accelerator experiments, as many as six will likely be needed for DUNE due to the expected detection of $\nu_\tau$'s \cite{Keloth:2017vdp,DeGouvea:2019kea,Machado:2020yxl,MammenAbraham:2022xoc}.
Since only three channels really need to be calculated, the remaining ones follow simply from unitarity.
For reactor experiments such as JUNO this may be slightly overkill and some computational time could be saved by removing a few simple lines of code, but the speed up from this is very marginal.

Second, we include all precalculations in the computation.
Some approaches in the literature precalculate certain parts of the probability that depend on the oscillation parameters but not the energy and then reuse those calculations for
a full calculation across an energy spectrum, see e.g.~\cite{Page:2023rpb,Arguelles:2021twb,Maltoni:2023cpv}.
Such an approach can lead to slight improvements in speed, but we find that the gain is marginal and it may require restructuring an existing analysis pipeline to take advantage of.
Nonetheless, if computing neutrino oscillation probabilities really needs to be speed up, our algorithm can benefit slightly by precalculating some expressions outside of the loop over energies.

The code can be found at \href{https://github.com/PeterDenton/NuFast}{github.com/PeterDenton/NuFast}.
The primary codes can be found in the \verb1c++1, \verb1f1, and \verb1py1 directories.
An additional \verb1Benchmarks1 directory contains the same c++ and fortran codes in a slightly restructured format, along with other codes, to reproduce the precision and speed benchmarks presented in the next section.

\subsection{c++}
The preferred method of running the code is with the c++ implementation, which is provided in the \verb1c++1 directory.
The primary function call is of the form
\lstset{language=c++}
\begin{lstlisting}
// NuFast.cpp
void Probability_Matter_LBL(double s12sq, double s13sq, double s23sq, double delta, double Dmsq21, double Dmsq31, double L, double E, double rho, double Ye, int N_Newton, double (*probs_returned)[3][3]);
\end{lstlisting}
The complex phase is taken in radians, the mass squared splittings are taken in eV$^2$, the baseline in km, the energy in GeV, the density $\rho$ in g/cm$^3$, and the electron fraction $Y_e$ is about 1/2 in the Earth.
The integer \verb1N_Newton1 can usually be taken to be zero; increasing this parameter increases the precision by many orders of magnitude per integer with a modest increase in run time.
Anti-neutrinos are handled with negative energies and the inverted mass ordering is handled with \verb!Dmsq31!$<0$.
The probabilities are returned as a $3\times3$ array with \verb!probs_returned[1][0]! containing $P_{\mu e}$, and so on.

An example function calling the code is provided in \verb1main()1.
Once the input variables are defined in the usual fashion, the function is called and read out as
\begin{lstlisting}
// NuFast.cpp
// Calculate the probabilities:
Probability_Matter_LBL(s12sq, s13sq, s23sq, delta, Dmsq21, Dmsq31, L, E, rho, Ye, N_Newton, &probs_returned);
// Print out the probabilities to terminal
for (int alpha = 0; alpha < 3; alpha++)
{
    for (int beta = 0; beta < 3; beta++)
    {
        printf("%d %d %g\n", alpha, beta, probs_returned[alpha][beta]);
    } // beta, 3
} // alpha, 3
\end{lstlisting}

The simple example code provided in the c++ directory can be compiled with the \verb1compile.sh1 script in the same c++ directory which creates the \verb1NuFast1 binary via the command:
\lstset{language=sh}
\begin{lstlisting}
# compile.sh
g++ -Ofast -ffast-math NuFast.cpp -o NuFast
\end{lstlisting}
The \verb1-Ofast1 and \verb1-ffast-math1 commands are not required.
They are fairly aggressive and may break IEEE standards, but do improve the performance somewhat.

A vacuum probability code is also provided which is the same algorithm, but without the $\lambda_3$ eigenvalue estimated in matter, and the input does not take \verb1rho1, \verb1Ye1, or \verb1N_Newton1.

\subsection{Fortran}
The inputs to the fortran code which can be found in the \verb1f1 directory are in the same units as the c++ code.
The main subroutine is of the form
\lstset{language=Fortran}
\begin{lstlisting}
! NuFast.f90
subroutine Probability_Matter_LBL(s12sq, s13sq, s23sq, delta, Dmsq21, Dmsq31, L, E, rho, Ye, N_Newton, probs_returned)
\end{lstlisting}
and an example calling of the subroutine is provided in the program NuFast in the same file.
After all the variables are defined in the usual fashion, the relevant portions are,
\begin{lstlisting}
! NuFast.f90
! Calculate the probabilities:
call Probability_Matter_LBL(s12sq, s13sq, s23sq, delta, Dmsq21, Dmsq31, L, E, rho, Ye, N_Newton, probs_returned)
! Print out the probabilities to terminal
do alpha = 1, 3
    do beta = 1, 3
        write (*,"(2I3,F10.6)") alpha, beta, probs_returned(alpha, beta)
    end do ! beta, 1, 3
end do ! alpha, 1, 3
\end{lstlisting}
The code is nominally written in \verb1f901, but can be easily adapted for other fortran versions as needed.
A module containing physics \verb1Parameters1 is also included, although this can also be adapted in a program's flow as needed.

Compilation is also straightforward by following the suggested compilation in the \verb1compile.sh1 script in the same \verb1f1 directory
\lstset{language=sh}
\begin{lstlisting}
# compile.sh
gfortran -fdefault-real-8 -Ofast -ffast-math NuFast.f90 -o NuFast
\end{lstlisting}
Again, the \verb1-Ofast1 and \verb1-ffast-math1 commands are not required and could conceivably cause unusual behavior, but have been tested and do improve computational performance somewhat.
The \verb1-fdefault-real-81 sets the default precision of \verb1real1 variables in the text to 8-byte double precision.
More (e.g.~16 for quadruple) or less (e.g.~4 for single) precision in the arithmetic can be easily adjusted with this compiler flag since the code does not specify the size of \verb1real1's in the code.

A vacuum version of the code is also provided, similarly to the c++ code.

\subsection{Python}
An additional python code is provided in the \verb1py1 directory which is a naive implementation in python.
Note that it is not optimized for speed and should be used only as a learning tool to understand the algorithm if python is one's preferred language.
If a python implementation of the algorithm is needed, we encourage the usage of \verb1F2py1, \verb1cython1, or other similar such translation tools from the optimized fortran or c++ codes.
Given the simplicity of the structure of the code, we have found that they translate to python with these tools in a fairly straightforward fashion.

The python function is given in the form
\lstset{language=python}
\begin{lstlisting}
# NuFast.py
def Probability_Matter_LBL(s12sq, s13sq, s23sq, delta, Dmsq21, Dmsq31, L, E, rho, Ye, N_Newton):
\end{lstlisting}
with inputs in the same form as the c++ and fortran codes.
The function returns a $3\times3$ array containing all nine probabilities.
The function can be called in the main program as

\begin{lstlisting}
# NuFast.py
# Calculate the probabilities:
probs_returned = Probability_Matter_LBL(s12sq, s13sq, s23sq, delta, Dmsq21, Dmsq31, L, E, rho, Ye, N_Newton)
# Print out the probabilities to terminal
for alpha in range(3):
    for beta in range(3):
        print(alpha, beta, probs_returned[alpha][beta])
\end{lstlisting}

\section{Performance}
\label{sec:performance}
There are multiple ways to quantify the performance of any code.
We focus on two relevant ones for the problem at hand, notably accuracy of the calculations and computational speed.

\subsection{Accuracy}
\label{sec:accuracy}
Two different useful definitions of accuracy are used in the literature: $\Delta P$ and $\Delta P/P$.
We focus on $\Delta P/P$, although we note that, at times, when $P\to0$, this can be somewhat misleading.
The reason why $\Delta P/P$ is usually the correct quantity to consider is because it is directly tied to what the experiments measure.

The number of events in any energy bin is $NP$, where N is the normalization and P is the relevant oscillation probability.
The statistical uncertainty on this is given by $\sqrt{NP}$.
Therefore for the error on P, $\Delta P$, so as to not effect the results must satisfy the following condition
\begin{align}
\sqrt{NP} &\gg N \Delta P = (NP) ~ \frac{\Delta P}{P} \,, \notag \\
 \Rightarrow \quad \frac{\Delta P}{P}  & \ll \frac{1}{\sqrt{NP}}\,.
 \end{align}
What this equation is telling us is that fractional error on P, must be much smaller than the inverse of the size of the statistical fluctuations in this bin.
Suppose the number of events in a given bin is  $10^{4}=NP$, a very large number for most neutrino experiments,
then the fractional uncertainty on $P$ should be
$$ \frac{\Delta P}{P} \ll 10^{-2}\,.$$
If the number of events per bin is smaller than $10^{4}$, then the allowable fractional error on $P$ can be larger than $10^{-2}$.
 
\begin{figure*}
\includegraphics[width=0.32\textwidth]{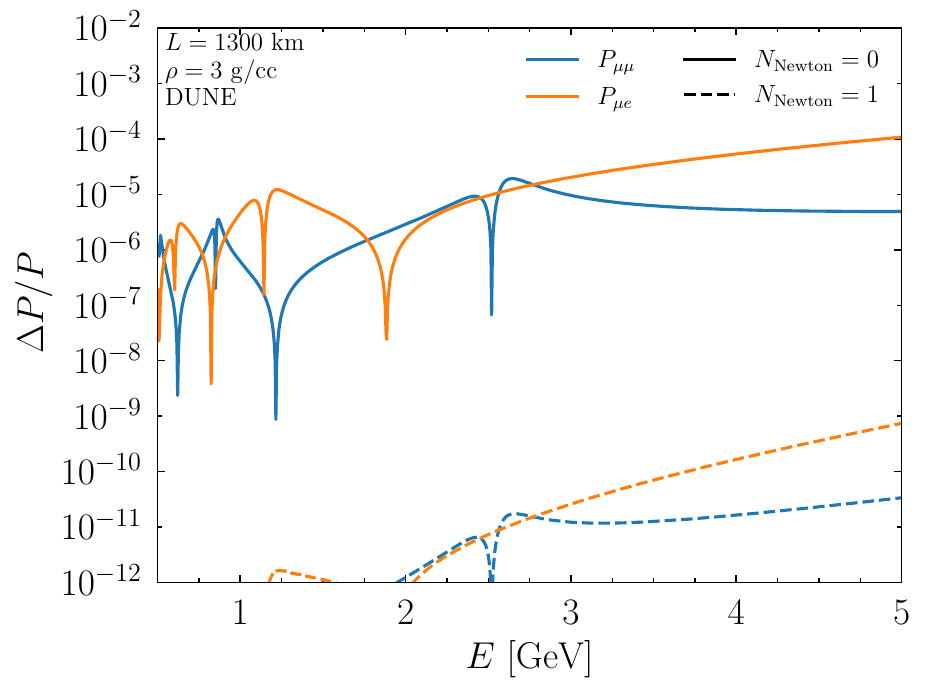}
\includegraphics[width=0.32\textwidth]{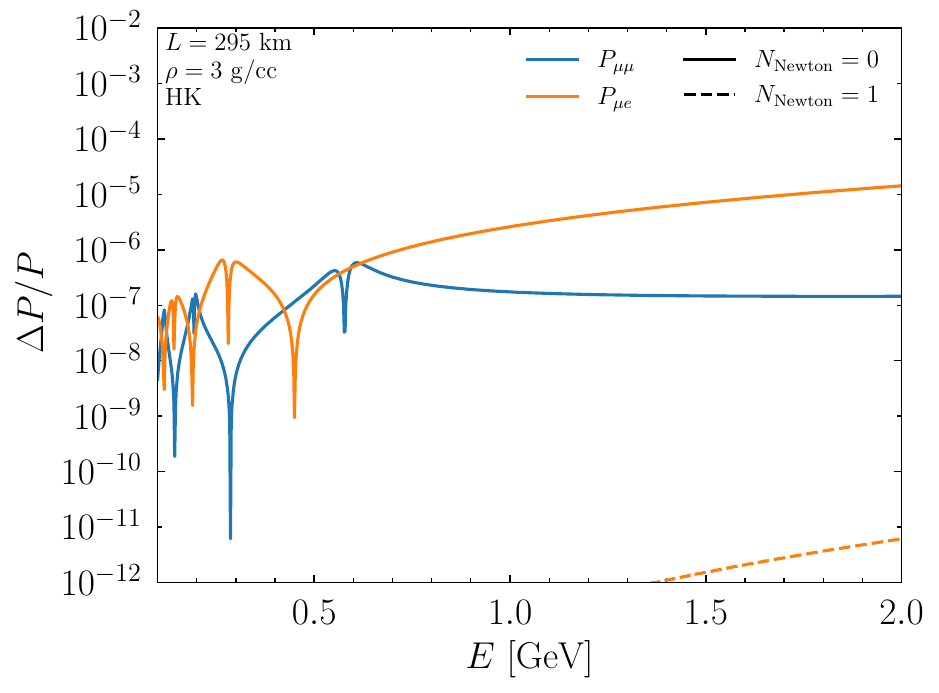}
\includegraphics[width=0.32\textwidth]{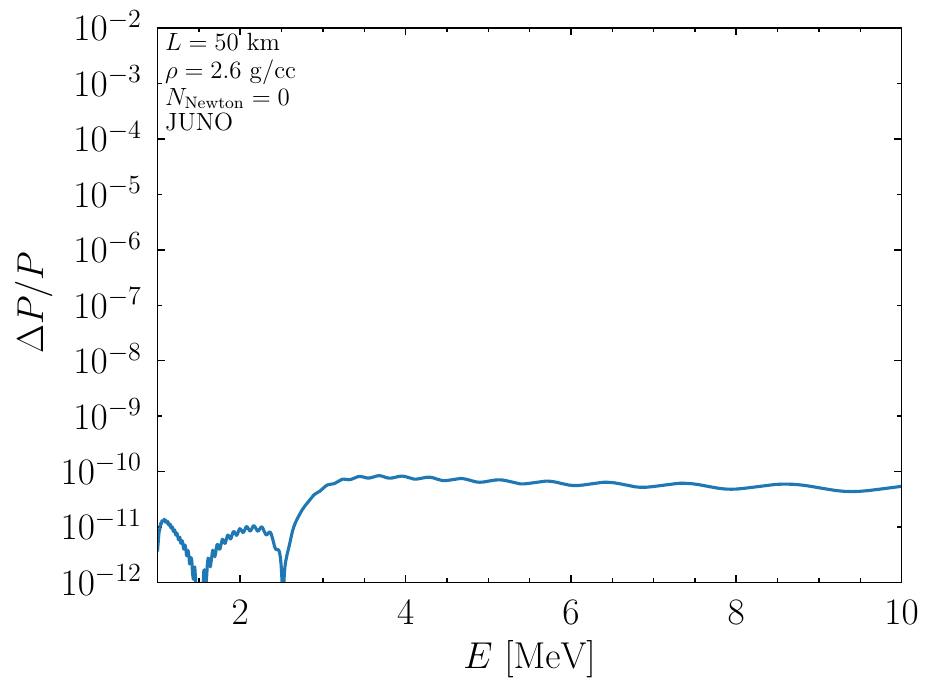}
\caption{The precision of the algorithm for three upcoming long-baseline configurations: DUNE (left), HK (center), and JUNO (right).
In the first two panels, the impact of the first NR correction is shown in the dashed curve providing $\sim5$ orders of magnitude of improvement.
A comparable result exists for reactors, but it is beyond the limit of double precision.}
\label{fig:precision}
\end{figure*}

In Fig.~\ref{fig:precision} we can see that, without any NR iterations, $\frac{\Delta P}{P} \le 10^{-4}$ in the energy range $0.5 < E/$GeV$<5$ and often much smaller, easily satisfying our requirement.
The largest fractional error is in the appearance channel at high energy where the oscillation probability, and thus the number of detected events, becomes very small.
With one NR iteration, the fractional precision improves by about five orders of magnitude, and even more performance gains can be found for additional iterations.
For the disappearance channel $ \frac{\Delta P}{P} \le 10^{-5}$ with no NR iterations and it converges even more rapidly as does the appearance channel.
The reason the disappearance channel converges more rapidly than the appearance channel, is because matter effects in the disappearance channel are very small even for DUNE, as discussed in \cite{Denton:2024thm}.
The conclusion here is that using zero NR iterations is most probably good enough for the current and under construction long baseline experiments, however,
since including one NR iteration costs only an additional $\sim10\%$, it may be worth the additional computational time.
 
\subsection{Speed}
\label{sec:speed}
For computational speed, we compute the oscillation probability many times and record the time per calculation.
We then repeat this many times to compute the standard deviation to account for any variations due to the behavior of the rest of the operating system.
The speed tests were done on an Intel \verb1i71 chip on a laptop.
Comparable results can be found on different chips with small changes to the optimization choices.

The compiler flags were varied between ``aggressive'' and ``conservative''.
Aggressive compiler optimizations including \verb1-Ofast1 for both the c++ and fortran code, which tends to include the most aggressive unrolling and calculations possible\footnote{In fact,
for the vacuum calculations, on some computers tested, aggressive compiler options would actually precompute all the oscillation probabilities.}.
We also include \verb1-ffast-math1 (which may already be included by \verb1-Ofast1).
This breaks IEEE standards, but performs calculations mathematically equivalent to the original code in ways that may see additional speed ups.
It is also known that \verb1-ffast-math1 breaks some other popular mathematics packages and can cause unwanted behavior, so care is required if this flag is included in a larger analysis chain.
Our ``conservative'' compiler flags are the default flags which are \verb1-O01.
While it is known that flags like \verb1-O31 are safe, reasonable, and likely faster than \verb1-O01, we limited our speed tests to these two cases for convenience.

We compared several different oscillation probability calculations to get a good handle on how our \verb1NuFast1 algorithm performs.
We first considered the vacuum probability which is calculated in the same way as described above, but with $\lambda_3$ just set to $\Delta m^2_{31}$.
We then compared our \verb1NuFast1 algorithm with no NR corrections and with 1 NR correction.
We have checked that additional NR corrections add linearly to the run-time.
We also used the exact cubic solution for $\lambda_3$ from \cite{cardano,Barger:1980tf,Zaglauer:1988gz}.
Note that our implementation of the exact oscillation probability is faster than that described in \cite{Zaglauer:1988gz}, the first complete description of the exact neutrino oscillation probabilities in matter,
due to both the overall structure of how the probability is calculated, how one eigenvalue is converted to all three, how all three are converted to the eigenvectors, and how the CP violating part is determined.
We then considered the algorithm in \cite{Page:2023rpb} (Page hereafter) which recently appeared\footnote{The
version of the code we used was the one without precomputing, see \href{https://github.com/Jamicus96/Nu-Pert-Compare/blob/compare_JP/src/JP.cpp}{github.com/Jamicus96/Nu-Pert-Compare/blob/compare\_JP/src/JP.cpp}.}.
Finally, we also consider the oscillation probability calculator used in GLoBES \cite{Huber:2004ka,Huber:2007ji}, specifically the algorithm developed in \cite{Kopp:2006wp},
as implemented in the latest version of GLoBES with everything stripped away from the full GLoBES software, except for the calculation of the probability.

Before getting to the results, we should stress that, while we are comparing the speed of our algorithm to others in the literature, our algorithm has a different goal than some other available codes.
For instance, there may be cases where precomputing certain trigonometric calculations outside of certain loops in an analysis pipeline may save some time; in which case our algorithm should be modified accordingly as well.
Our results can match the exact solution solution with a sufficient number of NR iterations for a given machine precision.
For example, for DUNE energies (and certainly also for lower energies such as HK or JUNO), with 2 NR iterations and double precision our results and the exact results match to machine precision. 
Here the precision is far greater than required for upcoming long-baseline neutrino oscillation experiments such as DUNE, HK, and JUNO.
Finally, the software suite GLoBES does much more than just compute neutrino oscillation probabilities and our code is not an attempt to recreate all of that,
rather it is designed to be dropped in to the appropriate place in standard three-flavor neutrino oscillation analysis pipelines.

We also note that this approach is not directly suitable for atmospheric or nighttime solar neutrinos, although some techniques may transfer over in a useful fashion.
In those areas there are a number of codes designed for these purposes such as NuSquids \cite{Arguelles:2021twb}, PEANUTS \cite{Gonzalo:2023mdh}, and others, each with individual pros and cons.

\begin{figure}
\centering
\includegraphics[width=\columnwidth]{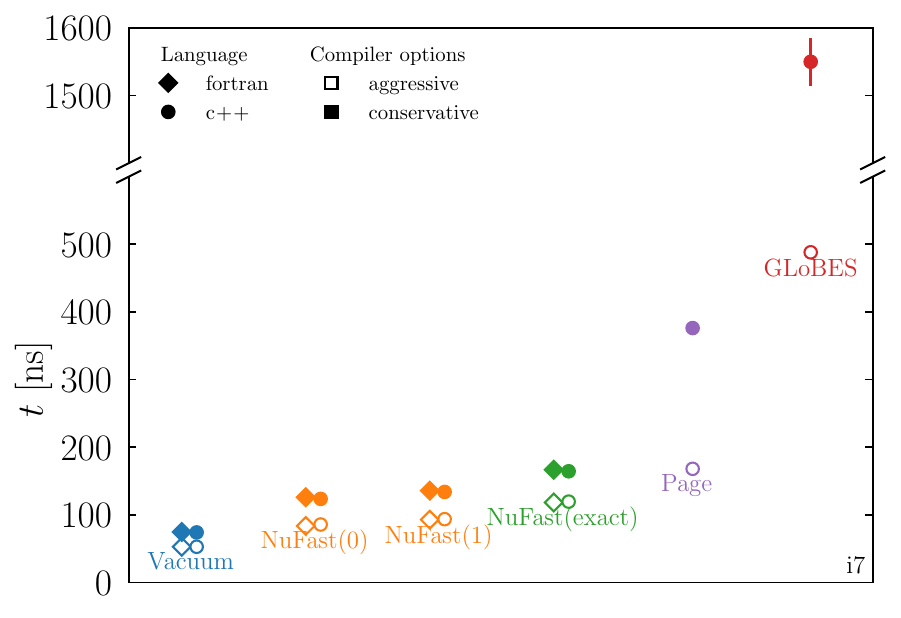}
\caption{We present our estimate of the computational time required to compute one set of nine neutrino oscillation probabilities on a single core on a laptop for different cases.
Each case is repeated for aggressive (\texttt{-Ofast} and \texttt{-ffast-math}) and conservative (default, i.e.~\texttt{-O0}) compiler flags.
The \texttt{NuFast} algorithm, presented in this paper, is coded up in both c++ and fortran.
The Page algorithm is that from \cite{Page:2023rpb} without any precomputations.
The algorithm used in GLoBES is actually coded in c.
There are vertical error bars on every point representing the uncertainty in the speed estimate.
Note the break in the vertical axis.}
\label{fig:speed}
\end{figure}

Finally, we present our numerical results in fig.~\ref{fig:speed}.
We see that the \verb1NuFast1 algorithm is not only faster than other approaches available, but is also close to as fast as the vacuum probability,
which can be reasonably thought of as fast as possible for neutrino oscillations in constant matter density.

\section{Conclusions}
\label{sec:conclusions}
Calculating neutrino oscillation probabilities, including the matter effect, is already becoming a computational issue, one that will only grow as experiments reach additional levels of precision necessitating additional Monte Carlo studies.
While a previous method \cite{Denton:2016wmg,Denton:2019ovn} by the same authors was found to lead to some $\mathcal O($few$)$ computational speed improvements in a real analysis code, the algorithm presented here,
dubbed \verb1NuFast1, performs better.
The speed up is due to several key innovations.

The first is the overall structure of how the oscillation probabilities are calculated.
Given one eigenvalue, ($\lambda_3$ performs best) the other two are derived from the properties of eigenvalues.
Then, using the adjugate approach \cite{Abdullahi:2022fkh} with the Eigenvector-Eigenvalue Identity (see e.g.~\cite{Denton:2019pka}) we compute the CP-even part of the probabilities.
We get the CP-odd part via the NHS identity \cite{Naumov:1991ju,Harrison:2002ee}.

The second is the means of efficiently approximating $\lambda_3$.
We start with the excellent approximation from DMP \cite{Denton:2016wmg} and get further improvements with Newton-Raphson (NR) corrections which provide $>5$ orders of magnitude improvement in the first iteration,
and the improvement is even faster -- quadratic -- thereafter.
With the methods of this paper, improving the lowest order approximation is numerically much simpler than in DMP as it just requires a simple NR iteration on $\lambda_3$, see eq.~\ref{eq:NR}.
Since this step is the only step with an approximation, additional NR iterations are guaranteed to take the entire probability to the exact probability.
We also note that no NR corrections are actually needed to provide probabilities significantly more precise than future experiments will need.

The third part is a careful implementation in the code.
Essentially regardless of the language, chip, or compiler flag, trigonometric functions, square roots, and so on are very expensive.
We have taken great care to minimize expensive computations.

The effect of this is an algorithm that is very precise (and can easily be made arbitrarily precise e.g.~beyond the double precision limit) and is faster than those available.
In fact, we believe that it is essentially as fast as possible, for the problem at hand.
We have provided easy to use, publicly available code in c++ and fortran for implementation in neutrino oscillation analysis pipelines \href{https://github.com/PeterDenton/NuFast}{github.com/PeterDenton/NuFast}.

\begin{acknowledgments}
PBD acknowledges support by the United States Department of Energy under Grant Contract No.~DE-SC0012704. SJP acknowledges support by the United States Department of Energy under Grant Contract No.~DE-AC02-07CH11359.  This project has also received support from the European Union's Horizon 2020 research and innovation programme under the Marie  Sklodowska-Curie grant agreement No 860881-HIDDeN as well as under the Marie Skłodowska-Curie Staff Exchange grant agreement No 101086085 - ASYMMETRY.

\end{acknowledgments}

\bibliography{Osc_Prob_Fast}

\appendix

\section{Optimal initial approximation for \texorpdfstring{$\lambda_3$}{lambda3}}
The initial approximation for $\lambda_3$ shown in eq.~\ref{eq:DMP+} is derived in a particular perturbative framework \cite{Minakata:2015gra,Denton:2016wmg} with a particular choice of initial basis.
This approximation is quite good and is fairly simple as it requires only diagonalizing a $2\times2$ matrix (solving a quadratic).

Here we investigate, numerically, if the specific parameters of the approximation used are optimal given a functional form of this simplicity.
To do so, we consider the more general form
\begin{align}
\lam{3}&={} \Delta m^2_{31} +\frac1{2}\Delta m^2_{yy}\left(x-1+\sqrt{1+x^2-2 x \cos 2\theta_{zz} } ~\right)\,,\notag \\
&\text{with}  ~ \Delta m^2_{yy} \equiv \Delta m^2_{ee}(1+y),  ~x\equiv \frac{a}{\Delta m^2_{yy}}\,, 
 \notag \\
&\text{and}~ \cos 2\theta_{zz}\equiv  (1+z) \cos 2\theta_{13}\,.
\end{align}
The choice used in this paper corresponds to $y=z=0$.

At the resonance energy, the contour crosses $z=0$ when $ y = - s^2_{12}c^2_{12}(\Delta m^2_{21}/\Delta m^2_{ee})^2 \approx -1.9 \times 10^{-4}$ as given in \cite{Parke:2020wha}.
In fact, as shown in fig.~\ref{fig:L3DMP}, we find that for a broad range of energies, the exact choice is very near to the approximate choice used here, at the $10^{-4}$ level for all energies.
Thus, without a more complicated functional form, one will not find a better approximation.

\onecolumngrid
\begin{center}
\includegraphics[width=0.45\textwidth]{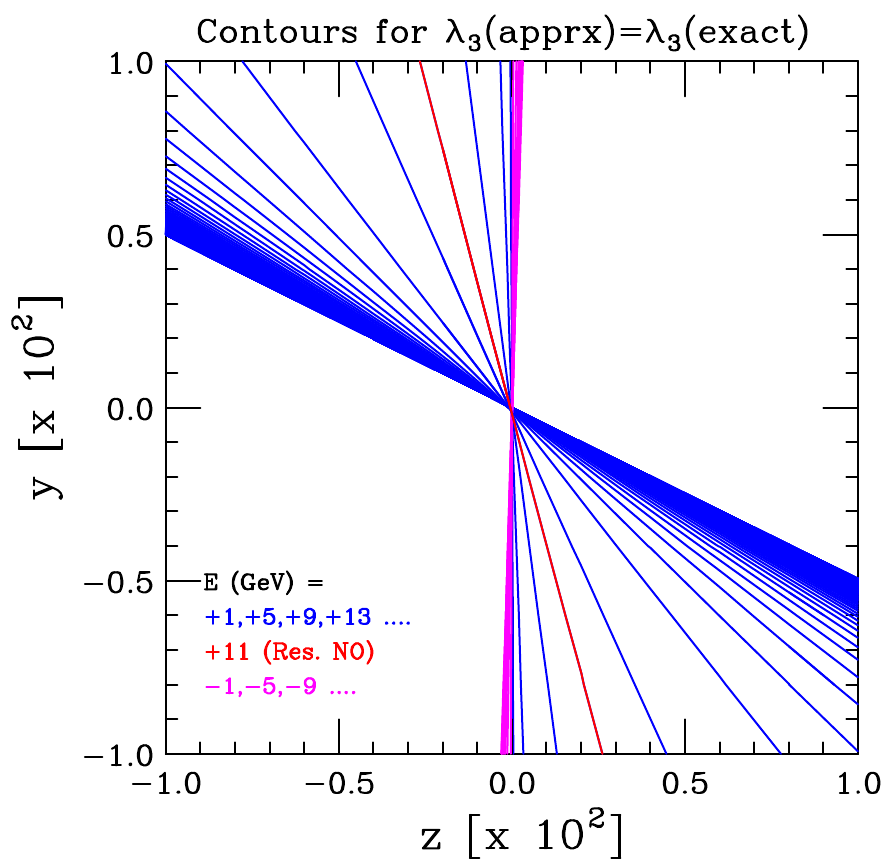}
\includegraphics[width=0.45\textwidth]{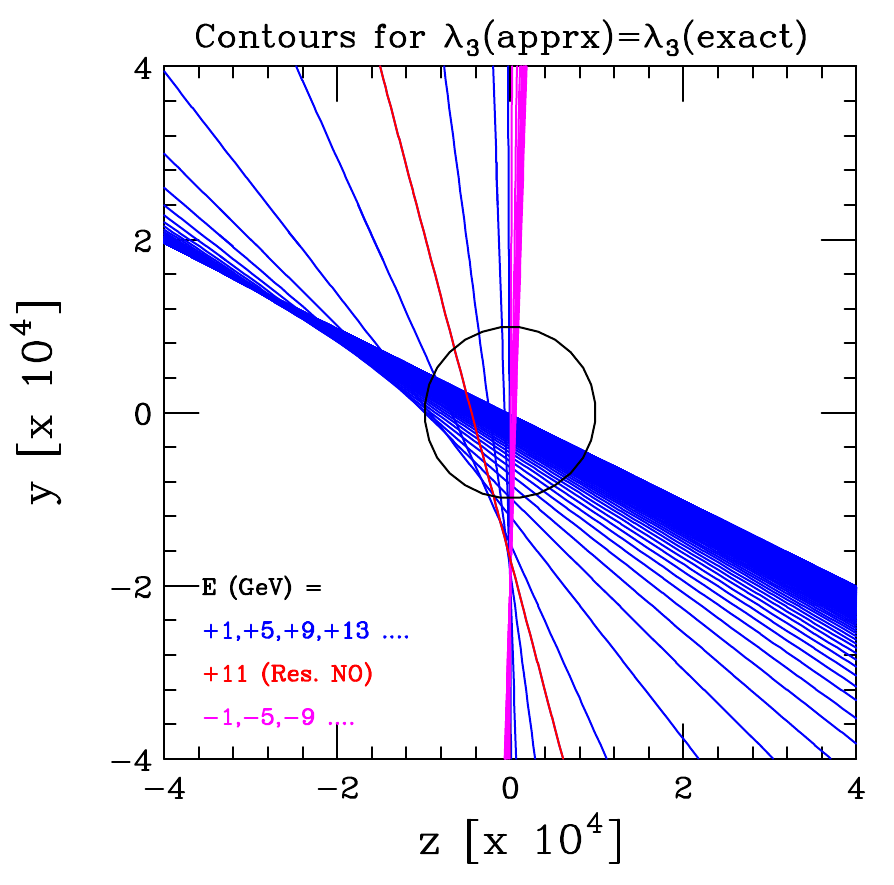}
\end{center}
\captionof{figure}{Contours in the $(z,y)$ plane where the approximate form for $\lambda_3$ is equal to exact value as one varies the energy.
The right panel is a zoomed in $25\times$ compared to the left panel.
The contours for all energies pass through the black circle centered at $(0,0)$ with  a radius $10^{-4}$.
The approximation in eq.~\ref{eq:DMP+} from \cite{Minakata:2015gra,Denton:2016wmg} is as good as you can do as without making the coefficients energy dependent.}
\label{fig:L3DMP}

\end{document}